\documentclass[aps,prl,twocolumn]{revtex4}
\usepackage{amssymb}
\usepackage{amsfonts}
\usepackage{amsmath}
\usepackage{graphicx}
\usepackage{epstopdf}
\begin{document}

\title{Superconducting properties in a candidate topological nodal line semimetal SnTaS$_2$ with a centrosymmetric crystal structure}

\begin{abstract}
We report the magnetization, electrical resistivity, specific heat measurements and band structure calculations of layered superconductor SnTaS$_2$. The experiments are performed on single crystals grown by chemical vapor transport method. The resistivity and magnetic susceptibility indicate that SnTaS$_2$ is a type-II superconductor with transition temperature $T_c = 3$ K. The upper critical field ($H_{c2}$) shows large anisotropy for magnetic field parallel to $ab$ plane ($H//ab$) and $c$ axis ($H//c$), and the temperature dependence of $H_{c2}$ for $H//ab$ shows obvious unconventional upward feature at low temperature. Band structure of SnTaS$_2$ shows several band crossings near the Fermi level, which form three nodal lines in the k$_z$ = 0 plane resulting in drumhead-like surface states when spin-orbit coupling is not considered. These results indicate that SnTaS$_2$ is a superconductor with possible topological nodal line semimetal character.

\end{abstract}

\author{Dong-Yun Chen$^{1,2,3}$, Yuelong Wu$^1$, Lei Jin$^4$, Yongkai Li$^{2,3}$, Xiaoxiong Wang$^1$, JunXi Duan$^{2,3}$,  Junfeng Han$^{2,3}$, Xiang Li$^{2,3}$, Yun-Ze Long$^1$, Xiaoming Zhang$^4$, Dong Chen$^{1,*}$, and Bing Teng$^{1,\dagger}$}

\affiliation{$^1$College of Physics, Qingdao University, Qingdao 266071, China}

\affiliation{$^2$Key Laboratory of Advanced Optoelectronic Quantum Architecture and Measurement, Ministry of Education, School of Physics, Beijing Institute of Technology, 100081, Beijing}

\affiliation{$^3$Micronano Centre, Beijing Key Lab of Nanophotonics and Ultrafine Optoelectronic Systems, Beijing Institute of Technology, Beijing, 100081, China}

\affiliation{$^4$School of Materials Science and Engineering, Hebei University of Technology, Tianjin 300130, China}

\maketitle

\section{Introduction}
Superconductors with nontrivial band structure have attracted much attention due to the possibility of realizing some novel quantum states, such as topological superconductors and Majorana fermions \cite{b1,b2}. Nontrivial band structure has been found in both topological insulators and topological semimetals \cite{b1,b3,b4,b5}. Topological semimetals can be further classified according to the configuration of band crossing near the Fermi level, including Dirac semimetal, Weyl semimetal, nodal line semimetal, etc \cite{b6,b7,b8,b9,b10,b11}. Superconductivity can be combined with the novel band structure by charge carrier doping or applying external pressure on topological materials. For example, topological insulator Bi$_2$Se$_3$ has been found superconducting when intercalated by Cu, Sr or Nb atoms \cite{b12,b13,b14}, or subjected to high pressure \cite{b15}. Dirac semimetal Cd$_3$As$_2$ and type-II Weyl semimetal WTe$_2$ can be induced as superconductors under high pressure \cite{b16,b17,b18}. Besides, some topological materials themselves are superconductors. The type-II Weyl semimetal MoTe$_2$ has a superconducting transition with $T_c=0.1$ K, and the $T_c$ can be dramatically enhanced to 8.2 K with high pressure \cite{b19}. PbTaSe$_2$ is found as a superconducting topological nodal line semimetal with a noncentrosymmetric structure \cite{b20,b21}. The later category of aforementioned superconductors does not suffer topological state shift from charge carrier doping or high pressure, providing an excellent playground for researching superconductivity with nontrivial band topology. 

The layered compound SnTaS$_2$, which is isoelectronic with PbTaSe$_2$, is another superconductor identified as early as 1973 \cite{b22}. Although the critical temperature for SnTaS$_2$ ($T_c=2.8$ K) has been reported \cite{b22,b23}, more detailed superconducting properties are still unknown. Moreover, we notice that SnTaS$_2$ has a centrosymmetric structure, which is different from that of PbTaSe$_2$. Whether the centrosymmetric SnTaS$_2$ can host the topological nontrivial band structure is also unknown. In this paper, we have systemically investigated the superconducting properties and the electronic structure of SnTaS$_2$. The magnetization, electric transport, and specific heat properties are studied in detail on the single crystal samples. We find large anisotropy in the upper critical field and coherence length. The temperature dependence of the upper critical field has an obvious upward feature for magnetic field parallel to $ab$ plane. Analysis of specific heat shows that SnTaS$_2$ is a moderately coupled superconductor. Using first principles calculations, we find that centrosymmetric SnTaS$_2$ exhibits a topological nodal line band structure when spin-orbit coupling (SOC) is not included. It features with three nodal lines centering the K point near the Fermi level, along with drumhead-like surface states corresponding to them. These properties are similar with those of PbTaSe$_2$. Our work suggests that SnTaS$_2$ is another system for investigating the novel properties of superconductors with topological nodal-line fermions.

\section{Experiment and Methods}
The single crystals of SnTaS$_2$ were grown by the chemical vapor transport  method with iodine as the transport agent. Polycrystalline Sn$_{0.33}$TaS$_2$ was synthesized previously by the solid state reaction of Sn, Ta and S powders in an evacuated quartz tube at 850 $^{\circ}$C. The obtained Sn$_{0.33}$TaS$_2$ polycrystals were mixed with Sn powders with element ratio Sn : Ta : S = 1.2 : 1 : 2 and sealed in an evacuated quartz tube together with iodine (3 mg/cm$^3$ in concentration). The excess Sn was used to restrain the appearance of Sn$_{0.33}$TaS$_2$. The quartz tube was then put into a two-zone furnace with 1000 $^{\circ}$C in hot zone and  970 $^{\circ}$C in cold zone for two weeks. Thin plate shaped single crystals were obtained with typical dimensions of 3 $\times$ 3 $\times$ 0.05 mm$^3$. The crystal structure of the obtained crytals was characterized by x-ray diffraction (XRD) on a Rigaku Smartlab x-ray diffractometer with Cu $K\alpha$ radiation at room temperature. The atomic ratio was determined by Oxford energy-dispersive x-ray (EDX) spectroscopy analysis. The magnetization, resistivity and specific heat of the samples were measured using a Quantum Design Physical Property Measurement System (PPMS). 

The band structure calculations were performed based on the density functional theory (DFT), as implemented in the Vienna $ab$ $initio$ simulation package (VASP) \cite{b24}. The generalized gradient approximation (GGA) with the realization of Perdew-Burke-Ernzerhof (PBE) functional was adopted for the exchange-correlation potential \cite{b25}. The cutoff energy was set to be 450 eV and the Brillouin zone was sampled with a 13$\times$13$\times$5 $\Gamma$-centered $k$-mesh. The energy convergence criterion was chosen as 10$ ^{-6}$ eV. The surface states were computed by using the Wannier$_-$tools package \cite{b26}.

\section{Results and Discussion}
\begin{figure}
\includegraphics[width=8.6cm]{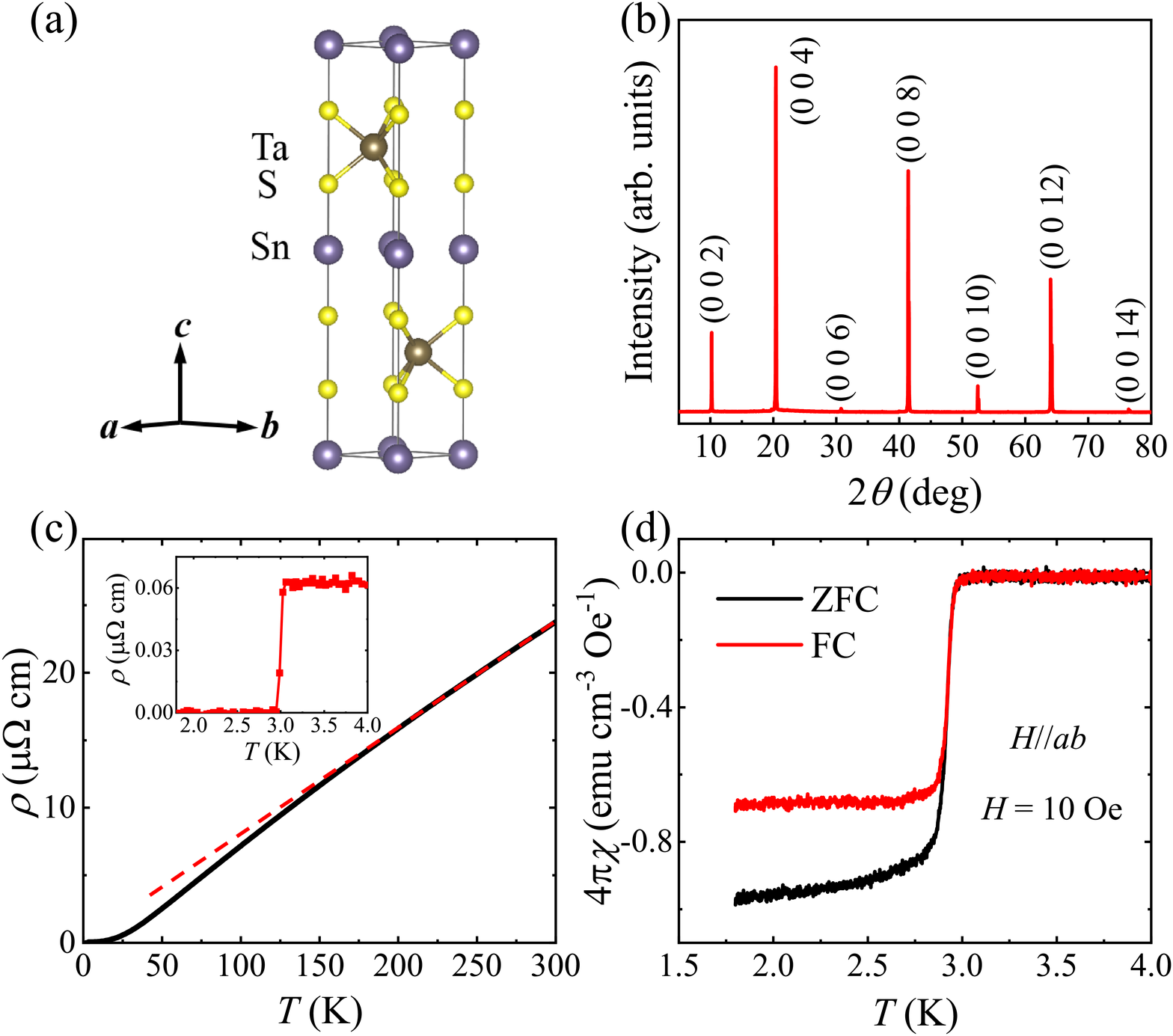}
\caption{\label{f1}(Color online)(a) Crystal structure of SnTaS$_2$. (b) The XRD pattern of SnTaS$_2$ single crystals with (00$l$) reflections.  (c) The temperature dependence of the resistivity with current flowing in the $ab$ plane. The straight dashed line is the guide to the linear behaviour of the resistivity in high temperature. The inset shows the superconducting transition of resistivity. (d) The zero-field-cooling (ZFC) and field-cooling (FC) magnetic susceptibility curves measured with magnetic field of 10 Oe and parallel to the $ab$ plane.}
\end{figure}

SnTaS$_2$ has a layered hexagonal structure (space group P6$_3$/mmc) with lattice parameters $a$ = $b$ = 3.309 \AA, and $c$ = 17.450 \AA. The layered structure is formed by the alternative stacking of TaS$_2$ and Sn layers, as shown in Fig. \ref{f1}(a). It should be noted that SnTaS$_2$ preserves the inversion and twofold screw rotation symmetries. Figure \ref{f1}(b) is the XRD pattern of the maximum surface of a plate-like single crystal. All of the peaks can be indexed as the (00$l$) reflections of SnTaS$_2$, indicating the single crystals perfectly oriented along the $c$-axis. The crystals can be further confirmed as SnTaS$_2$ by powder XRD and EDX measurements.

Figure \ref{f1}(c) displays the typical temperature dependence of the resistivity for SnTaS$_2$ single crystals with current applied in the $ab$ plane. The resistivity shows a metallic behavior and the residual resistivity ratio RRR = $\rho$(300 K)/$\rho$(4 K) is as high as 380, indicating the high quality of the samples. The linear temperature dependence of the resistivity at high temperature suggests the dominance of electron-phonon scattering. As shown in the inset of Fig.  \ref{f1}(c), the resistivity has a sharp superconducting transition, and the critical temperature ($T_c$) can be determined as 3.0 K using the criterion of 50\% point on the transition curve. To further demonstrate the superconductivity of the samples, we measured the dc susceptibility in the zero-field-cooling (ZFC) and field-cooling (FC) processes with magnetic field of 10 Oe parallel to the $ab$ plane, as displayed in Fig.  \ref{f1}(d).  The demagnetization effect is not considered due to the thin plate shaped sample with magnetic field parallel to the sample plane. The $T_c$ determined from the susceptibility curve is about 2.97 K, close to that determined from resistivity data.  The superconducting shielding volume fraction at 1.8 K is close to 100\%. In addition, the FC curve is not coincident with the ZFC curve, suggesting a type-II superconductor. However, the FC curve is not much larger than the ZFC curve, indicating few vortices pinning. 

\begin{figure}
	\includegraphics[width=8.6cm]{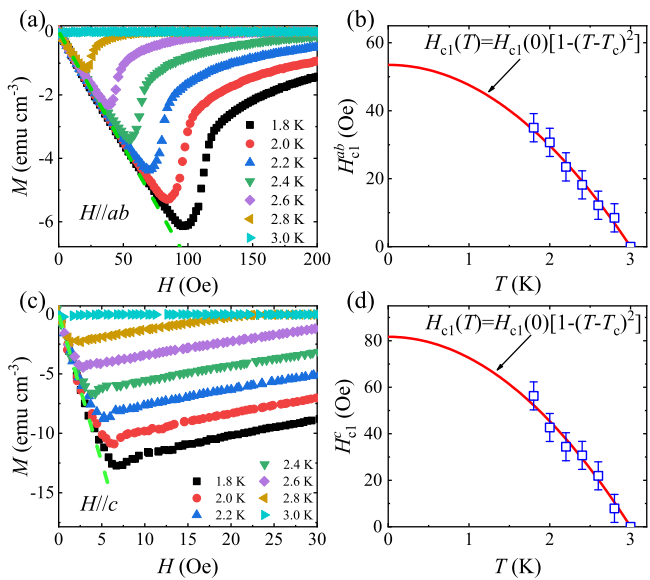}
	\caption{\label{f2}(Color online) (a) (c)The zero-field cooled magnetization $M(H)$ curves at different temperatures with magnetic field parallel to the $ab$ plane and $c$ axis, respectively. The straight dashed lines are the linear fit to the 2 K curves in low-field range. (b) (d) The lower critical field for magnetic field parallel to the $ab$ plane and $c$ axis, respectively, with the fit lines using equation $H_{c1}(T)=H_{c1}(0)[1-(T/T_c)^2]$. The demagnetizaion correction is considered in $H_{c1}^{c}$.}
\end{figure}

To estimate the anisotropic lower critical field ($H_{c1}$),  the zero-field cooled magnetization $M(H)$ curves were measured with H//$ab$ and H//$c$, as shown in Fig.  \ref{f2}(a) and \ref{f2}(c), respectively. The values of $H_{c1}$ are determined as the points for $M(H)$ deviating 5\% from the linear fitted curves. To get the accurate values of $H_{c1}$, the demagnetization effect should be considered. For a perfectly diamagnetic superconductor, the magnetic field lines are excluded from the inside of the sample and have higher density in the outside. This makes a higher field that the sample feels around it and a more pronounced diamagnetization slope, i. e. $M/H_a = -1/(1-N)$, where $N$ is the demagnetization factor. For the thin plate shaped sample, $N$ is almost 0 for magnetic field parallel to the sample's surface and nearly 1 when magnetic field is perpendicular to the surface. Thus, for the thin SnTaS$_2$ plates, the $H_{c1}$ for $H//ab$ ($H_{c1}^{ab}$) can be determined directly from the magnetization curves without demagnetization correction, as shown in Fig. \ref{f2}(b). By contrast, the demagnetization correction can not be neglected when we determine the $H_{c1}$ for $H//c$ ($H_{c1}^{c}$). To calculate the demagnetization factor $N$, we employ the relation \cite{b27} 
\begin{equation}
N=1-\left.1\middle/\left(1+q_{disk}\frac{a}{c}\right) \right.,
\end{equation}
\begin{equation}
q_{disk}=\frac{4}{3\pi}+\frac{2}{3\pi}\tanh\left[1.27\frac{c}{a}\ln\left(1+\frac{a}{c}\right)\right],
\end{equation}
where $a$ and $c$ are the dimension perpendicular to the magnetic field and the thickness of the sample, respectively. Here, the demagnetization correction has been considered in determining the values of $H_{c1}^c$ in Fig. \ref{f2}(d) with $N \approx 0.9473$. Both of the $H_{c1}^c$ and $H_{c1}^{ab}$ data can be fitted using $H_{c1}(T)=H_{c1}(0)[1-(T/T_c)^2]$. The lower critical fields at zero temperature for the two directions are $H_{c1}^{ab}(0) = 53.5$ Oe and $H_{c1}^c(0)=81.7$ Oe, respectively. 

\begin{figure}
	\includegraphics[width=8.6cm]{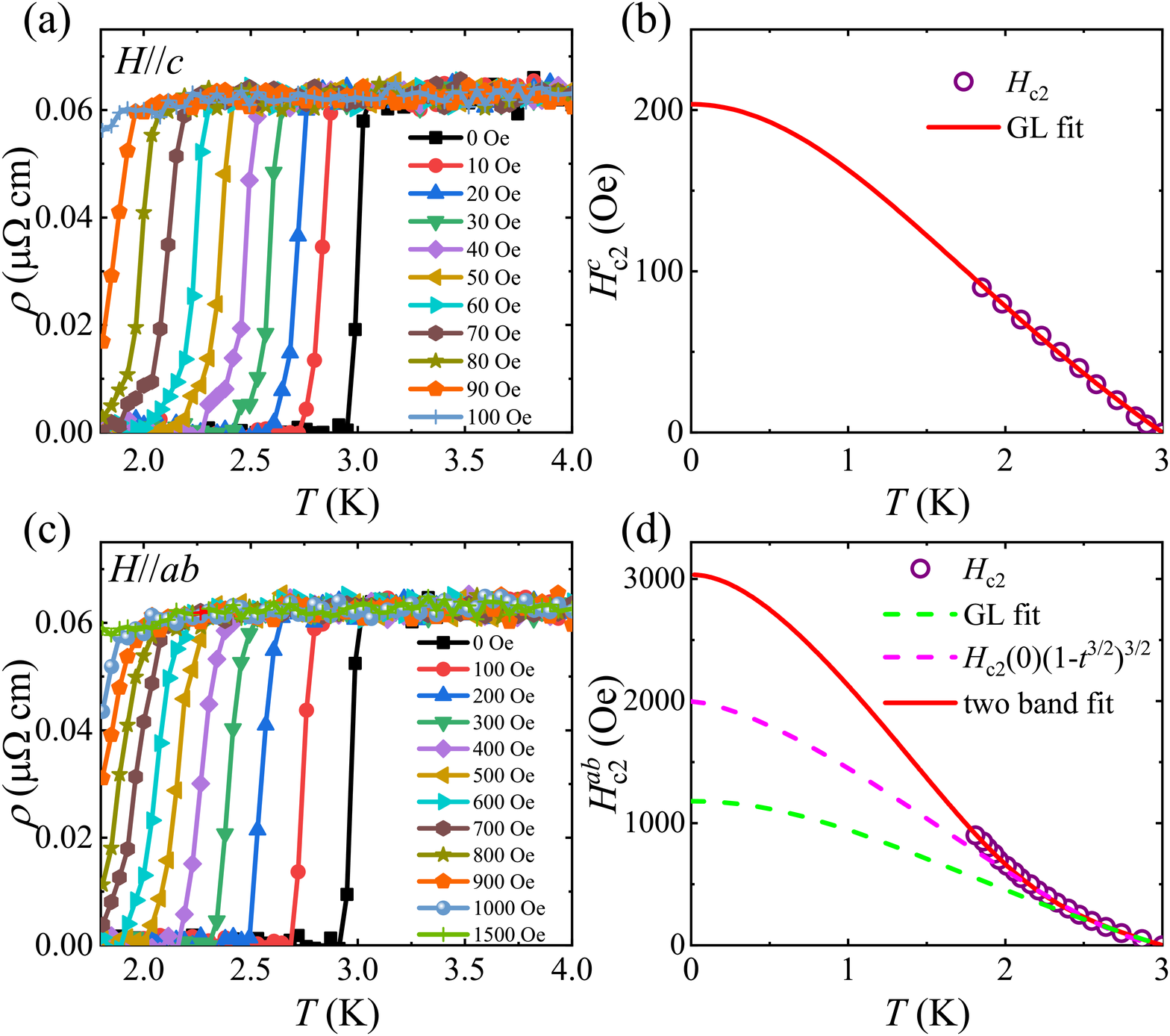}
	\caption{\label{f3}(Color online) (a) (c) The low temperature $\rho(T)$ curves under different magnetic fields for $H//c$ and $H//ab$, respectively. (b) The temperature dependence of the upper critical field extracted from the resistivity curves with $H//c$. The data are fitted by GL equation. (d) The temperature dependence of the upper critical field extracted from the resistivity curves with $H//ab$. The data are fitted by three different equations.}
\end{figure}

The low temperature resistivity under various magnetic fields with $H//c$ and $H//ab$ are presented in Fig. \ref{f3}(a) and \ref{f3}(c), respectively. The superconducting transition is very sharp for the zero-field curve and is broaden lightly by the applied fields. With the 50\% criterion to determine the transition temperatures, the upper critical fields for $H//c$ ($H_{c2}^{c}$) and $H//ab$ ($H_{c2}^{ab}$) as the functions of temperature are given in Fig. \ref{f3}(b) and \ref{f3}(d), respectively. The temperature dependence of $H_{c2}^{c}$ can be well fitted with Ginzberg-Landau (GL) equation $H_{c2}(T)=H_{c2}(0)(1-t^2)/(1+t^2)$, where $t=T/T_c$. The upper critical field at $T=0$ K for $H//c$ is accordingly estimated as $H_{c2}^{c}(0) = 203.6$ Oe. For $H//ab$, the temperature dependence of $H_{c2}^{ab}$ has an obvious upward feature and deviates from the GL equation for $T < 2.5$ K. This upward feature for $H_{c2}^{ab}(T)$ is also found in PbTaSe$_2$ \cite{b20,b28,b29,b30}, where the upper critical field can be roughly fitted by equation $H_{c2}(T)=H_{c2}(0)(1-t^{3/2})^{3/2}$. However, the fitting curve using this formula is also lower than the $H_{c2}^{ab}(T)$ data in low temperature region (Fig. \ref{f3}(d)), indicating the upward feature in SnTaS$_2$ is more obvious than that of PbTaSe$_2$. 

The enhancement of $H_{c2}$ in low temperature has several possible origins, such as: (1) dimensional crossover \cite{b31}, (2) presence of impurities and disorder \cite{b32,b33}, (3) multiband effect \cite{b34}, (4) melting of the vortex lattice associate with quantum critical point \cite{b35}, and (5) the non-local effect in the clean limit \cite{b36}. The dimensional crossover means that the superconductor changes from bulk superconductor to an stacked array of two-dimensional superconducting layers with the coherence length perpendicular to the layers ($\xi_c$) getting smaller than the distance between adjacent layers with decreasing temperature. This theory has been used to explain the upward curvature in $H_{c2}^{ab}(T)$ of organic molecules intercalated 2$H$-TaS$_2$ \cite{b37,b38}, in which the $H_{c2}^{ab}$s can exceed the Pauli paramagnetic limit at moderate temperatures and the coherence lengths are comparable with the TaS$_2$ layer distance. In our case, the $H_{c2}^{ab}$ is much smaller than the Pauli paramagnetic limiting field ($\mu_0H_P=5.5$ T) down to 1.8 K, and the $\xi_c$ is much larger than the layer distance (see below). These facts demonstrate that the layers in SnTaS$_2$ are not decoupled and it is a bulk superconductor. On the other hand, the high value of RRR = 380 indicates the high quality and low density of defects in our samples. This excludes the possibility of the scattering by impurities and supports the non-local effect scenario. The current properties of SnTaS$_2$ show no trace for the quantum critical point. The result of DFT computations discussed later shows several bands cross the Fermi level. Therefore, the multiband effect and the non-local effect are the most likely origins for the upward curvature of $H_{c2}^{ab}(T)$. 

\begin{table}
	\caption{The superconducting parameters of SnTaS$_2$ single crystal.}
	\label{t1}
	\setlength{\tabcolsep}{5mm}
	\begin{tabular}{lccc}
		\hline\hline
		&$H//ab$&$H//c$&\\
		\hline
		$H_{\rm c1}(0)$ (Oe)&53.5&81.7&\\
		$H_{\rm c2}(0)$ (Oe)&3034&203.6&\\
		$\xi(0)$ (nm)&127&8.5&\\
		$\kappa(0)$&7.58&5.07&\\
		$\lambda(0)$ (nm)&64.4&962.4&\\
		$\gamma_{\rm anis}(0)$& & &14.9\\
		$\Delta C/\gamma T_{\rm c}$& & &1.23\\
		$\Theta_{\rm D}$ (K)& & &154.4\\
		$\lambda_{\rm ep}$& & &0.66\\
		\hline\hline
	\end{tabular}
\end{table}

To estimate $H_{c2}^{ab}(0)$, we try to fit the $H_{c2}^{ab}(T)$ data using the equation for a two-band superconductor \cite{b34}, i. e. 
\begin{equation}
\begin{split}
a_0[\ln t+U(t)][\ln t+U(\eta h)]+a_1[\ln t+U(h)]\\
+a_2[\ln t+U(\eta h)]=0,
\end{split}
\end{equation}
where $t=T/T_c$, $h=\frac{H_{c2}D_1}{2\phi_0T}$, $\eta=\frac{D_2}{D_1}$, $U(x)=\psi(x+\frac{1}{2})-\psi(\frac{1}{2})$, $a_0=2(\lambda_{11}\lambda_{22}-\lambda_{12}\lambda_{21})/\lambda_0$, $a_1=1+(\lambda_{11}-\lambda_{22})/\lambda_0$, $a_2=1-(\lambda_{11}-\lambda_{22})/\lambda_0$, and $\lambda_0=\sqrt{(\lambda_{11}-\lambda_{22})^2+4\lambda_{12}\lambda_{21}}$. $\psi(x)$ is the digamma function. $D_1$ and $D_2$ are the intraband diffusivities of each band. $\lambda_{11}$ and $\lambda_{22}$ are the intraband coupling constants, whereas $\lambda_{12}$ and $\lambda_{21}$ are the interband coupling constants. As shown in Fig. \ref{f3}(d), the experimental data can be well fitted by the two-band formula, which gives $H_{c2}^{ab}(0)=3034$ Oe. The $H_{c2}^{ab}(0)$ is much lower than the Pauli limiting field, suggesting the upper critical field is limited by the orbital effect. The anisotropic ratio of $H_{c2}$, $\gamma_{\rm anis}(0) = H_{c2}^{ab}(0)/H_{c2}^{c}(0)$, is as large as 14.9, larger than those of 2$H$-TaS$_2$ ($\gamma_{\rm anis}$ = 6) and  PbTaSe$_2$ ($\gamma_{\rm anis}$ = 11.6) \cite{b29,b38}.

Based on the GL theory, the anisotropic coherence length is given by $H_{c2}^{ab}=\Phi_0/(2\pi\xi_{ab}\xi_c)$ and $H_{c2}^{c}=\Phi_0/(2\pi\xi_{ab}^2)$, where $\Phi_0$ is the flux quantum \cite{b39}. Accordingly, the anisotropic GL coherence length at zero temperature can be determined as $\xi_{ab}(0)=127$ nm and $\xi_{c}(0)=8.5$ nm, and the anisotropic ratio is $\xi_{ab}/\xi_{c}=14.9$. As mentioned above, the $\xi_c$ is much larger than the TaS$_2$ layer distance ($\sim 8.7$ \AA), indicating the bulk superconductivity of this system. The GL parameter $\kappa_i(0)$ along the $i$ direction can be obtained by the equation $H_{c2}^i(0)/H_{c1}^i(0)=2\kappa_i^2(0)/{\rm ln}\kappa_i(0)$. The GL parameter $\kappa_i(0)$ is related with the anisotropic GL penetration length $\lambda_i(0)$ and coherence length $\xi_i(0)$ by equations $\kappa_c(0)=\lambda_{ab}(0)/\xi_{ab}(0)$ and $\kappa_{ab}(0)=\lambda_{ab}(0)/\xi_{c}(0)=[\lambda_{ab}(0)\lambda_{c}(0)/\xi_{ab}(0)\xi_c(0)]^{1/2}$, in which $\xi_{ab}/\xi_c=\lambda_{c}/\lambda_{ab}$. With these relations, the GL parameter $\kappa_i(0)$ and the anisotropic $\lambda_i(0)$ can be determined, as listed in Table \ref{t1}. 

\begin{figure}
	\includegraphics[width=8.6cm]{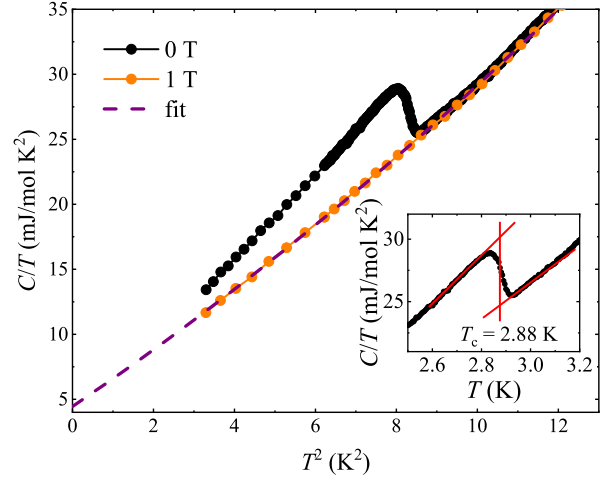}
	\caption{\label{f4}(Color online) The specific heat divided by temperature ($C/T$) as the function of $T^2$ with $\mu_0H=0$ and 1 T. The curve for $\mu_0H=1$ T is fitted by relation $C/T=\gamma+\beta T^2+\delta T^4$. The inset shows the isoentropic method for $T_c$ determination.}
\end{figure}

To further investigate the superconducting properties of SnTaS$_2$, we performed specific heat measurements and analysis. The low temperature specific heat under the fields of $\mu_0H$ = 0 and 1 T are demonstrated as the relationships of $C/T$ versus $T^2$ in Fig. \ref{f4}. With $\mu_0H=0$ T, the specific heat shows a sharp jump at $T_c=2.88$ K, which is determined by the isoentropic method shown in the inset. The superconducting transition is completely suppressed by magnetic field of 1 T, and the $C/T-T^2$ curve under $\mu_0H=1$ T can be well fitted by the formula $C/T=\gamma+\beta T^2+\delta T^4$. The fit yields the normal state Sommerfeld coefficient $\gamma=4.45$ mJ/mol K$^2$, and the phonon specific coefficient $\beta=2.11$ mJ/mol K$^4$. The value of $\Delta C/\gamma T_c$ is estimated as 1.23, which is smaller than the value of BCS theory (1.43). With the formula $\Theta_D=[(12/5\beta)\pi^4nN_Ak_B]^{1/3}$, where $n=4$ for SnTaS$_2$ and $N_A$ is the Avogadro constant, the Debye temperature is estimated as $\Theta_D=154.4$ K. The electron-phonon coupling constant $\lambda_{ep}$ can be calculated using the McMillans Formula \cite{b40}: 
\begin{equation}
\lambda_{ep}=\frac{1.04+\mu^*{\rm ln}(\Theta_D/1.45T_c)}{(1-0.62\mu^*){\rm ln}(\Theta_D/1.45T_c)-1.04}.
\end{equation}
The value of $\lambda_{ep}$ is 0.66 assuming $\mu^*=0.13$. This value is smaller than 1.0, the minimum value of strong coupling, indicating SnTaS$_2$ a moderately coupled superconductor. Both of the values of $\Theta_D$ and $\lambda_{ep}$ are close to those of PbTaSe$_2$ \cite{b20,b28,b29}. With these results, the noninteracting density of states at Fermi level can be calculated by $N(E_F)=3\gamma/[\pi^2k_B^2(1+\lambda_{ep})]$, which gives $N(E_F)=1.14$ states eV$^{-1}$ per formula unit. It can be noticed that the above parameters are different from those in the previous work \cite{b23}. This discrepancy may be due to the different sample quality caused by the changed growth conditions. 

\begin{figure}
	\includegraphics[width=8.6cm]{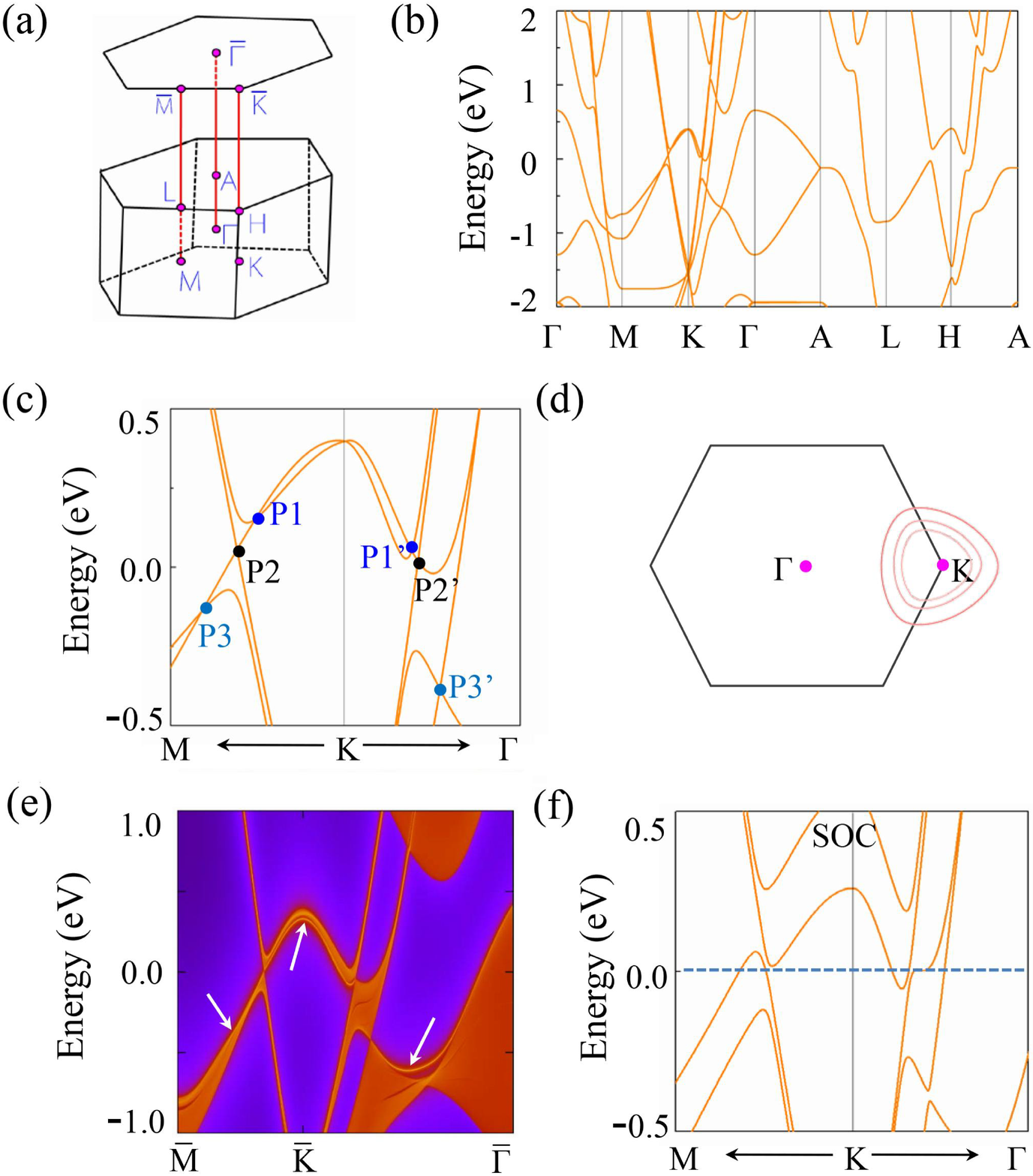}
	\caption{\label{f5}(Color online) (a) Bulk and (001)-surface Brillouin zone of  SnTaS$_2$. (b) Electronic band structure of SnTaS$_2$ without SOC. (c) Enlarged band structure along M-K and K-$\Gamma$ paths. (d) Illustration of the three nodal lines centering the K point in the k$_z$ = 0 plane. (e) The Sn-terminated (001) surface band structure for SnTaS$_2$. The drumhead-like surface states are pointed by white arrows. (f) Electronic band structure of SnTaS$_2$ with SOC included.}
\end{figure}

In view of the similarity in element component of SnTaS$_2$ with the topological nodal line semimetal PbTaSe$_2$, we study the band structure of SnTaS$_2$ through the first principles calculations. Figure \ref{f5}(a) is the schematic diagram for bulk and (001)-surface Brillouin zone of SnTaS$_2$. Figure \ref{f5}(b) clearly manifests a metallic band structure with several bands crossing the Fermi level. As shown by the enlarged band structure in Fig. \ref{f5}(c), there exist six band crossing points at M-K and K-$\Gamma$ paths near the Fermi level without SOC in account. By performing more careful calculations on the band structure nearby, we find these band crossing points are not isolated but belong to three nodal lines centering the K point in the k$_z$ = 0 plane, as shown in Fig. \ref{f5}(d). This indicates SnTaS$_2$ is a topological nodal line semimetal in the absence of SOC. In addition, the drumhead-like surface states from the nodal lines are quite visible, as pointed by the arrows in Fig. \ref{f5}(e). When SOC is included, the nodal lines in SnTaS$_2$ are gapped, as shown in Fig. \ref{f5}(f). The sizes of SOC gaps in SnTaS$_2$ are in the range of 15-180 meV, which are comparable with typical nodal line materials such as TiB$_2$, Cu$_3$PdN, Mg$_3$Bi$_2$, and CaAgAs \cite{b41,b42,b43,b44,b45}. From Fig. \ref{f5}(b) and (f), we can observe several electron-like and hole-like bands cross the Fermi level, consistent with the multiband scenario which is a possible origin for the upward curvature of $H_{c2}^{ab}(T)$ in our samples. The density of states at Fermi level calculated from the band structure is 1.04 states eV$^{-1}$ per formula unit with SOC, in good agreement with the value obtained from specific heat measurements. 

It is interesting that, similar with PbTaSe$_2$,  SnTaS$_2$ is also a superconducting nodal line semimetal, which is isoelectronic but not isostructural with PbTaSe$_2$. Although they have the similar electronic structure, the nodal lines in these two materials come from different origins. The nodal lines in noncentrosymmetric PbTaSe$_2$ are protected by the mirror reflection symmetry \cite{b21}, while the nodal lines in SnTaS$_2$ are protected by time-reversal and inversion symmetries. The different symmetries cause that, when SOC is considered, the nodal lines are persistent in PbTaSe$_2$ but gapped in SnTaS$_2$ \cite{b21}. These common and different properties with PbTaSe$_2$ make SnTaS$_2$ as a great platform to further investigate the properties of superconducting nodal line semimetals.

\section{Conclusion}
In summary, we report the magnetic, transport, specific heat properties and electronic structure of centrosymmetric compound SnTaS$_2$, which is a layered type-II superconductor. Large anisotropy is found in the upper critical field and GL coherence length. An obvious upward curvature is observed in the upper critical field curve for $H//ab$, maybe due to the multiband effect or the non-local effect, which needs further investigations to clarify. The electron-phonon coupling constant is determined as 0.66, indicating a moderately coupled superconductor. The band structure of SnTaS$_2$ exhibits three nodal lines in the k$_z$ = 0 plane near the Fermi level with drumhead-like surface states. With the similar crystal structure but different symmetries with noncentrosymmetric PbTaSe$_2$, SnTaS$_2$ is considered as a promising system to research the novel properties of superconducting topological nodal line semimetals.

\section{Acknowledgements}
This work was supported by the National Natural Science Foundation of China (Grant No. 11804176, 11734003), the National Key Research and Development Program of China (Grant No.  2016YFA0300604), Shandong Provincial Natural Science Foundation, China (Grant No. ZR2018BA030), and China Postdoctoral Science Foundation (Grant No. 2018M632609).

$*$dchen@qdu.edu.cn

$\dagger$5108tb@163.com

\end{document}